\documentclass[aps,reprint,twocolumn,showpacs,superscriptaddress,pra]{revtex4-1}
\usepackage{amsmath,amssymb,amsfonts}
\usepackage[pdftex]{graphicx}
\usepackage{microtype}
\usepackage{color}

\begin{document}

\title{Exciton dynamics in emergent Rydberg lattices}

\newcommand{\Nottingham}{School of Physics and Astronomy,
  The University of Nottingham, University Park,
  NG7 2RD Nottingham, United Kingdom}
\newcommand{\Durham}{Joint Quantum Centre (JQC) Durham-Newcastle,
   Department of Physics, Durham University,
   Durham, DH1 3LE, UK}

\author{S. Bettelli}          \affiliation{\Nottingham}
\author{D. Maxwell}           \affiliation{\Durham}
\author{T. Fernholz}          \affiliation{\Nottingham}
\author{C. S. Adams}          \affiliation{\Durham}
\author{I. Lesanovsky}        \affiliation{\Nottingham}
\author{C. Ates}              \affiliation{\Nottingham}

\date{\today}

\begin{abstract}
The dynamics of excitons in a one-dimensional ensemble with partial spatial
order are studied. During optical excitation, cold Rydberg atoms
spontaneously organize into regular spatial arrangements due to their mutual
interactions. This emergent lattice is used as the starting point to study
resonant energy transfer triggered by driving a $nS$ to $n^\prime P$
transition using a microwave field. The dynamics are probed by detecting the
survival probability of atoms in the $nS$ Rydberg state. Experimental data
qualitatively agree with our theoretical predictions including the mapping
onto XXZ spin model in the strong-driving limit. Our results suggest that
emergent Rydberg lattices provide an ideal platform to study coherent energy
transfer in structured media without the need for externally imposed
potentials.
\end{abstract}

\maketitle

\section{Introduction}\label{sec:intro}
The investigation of far-from-equilibrium phenomena overarches the fields of
physics, chemistry and biology. Systems out of equilibrium feature
fascinating phenomena such as the formation of complex ordered structures in
spite of a rather simple underlying microscopic description \cite{crho:93}.
Moreover, dynamical phenomena such as resonant energy transfer are of
central practical importance as they underlie many fundamental physical
processes in molecular aggregates \cite{sc:03}, photosynthesis
\cite{enca+:07} and novel materials such as organic solar cells
\cite{scru:06}.

Gases of cold atoms have been established in the past years as a platform
which permits the detailed investigation of systems in and out of
equilibrium \cite{mobr+:12,grku+:12,trch+:12}. Recently atoms excited to
high-lying Rydberg states have become a major focus due to their strong
interactions over large distances (several micrometers) and short time
scales (nanoseconds) allowing to address fundamental questions in many-body
physics. On the one hand the strong interactions give rise to the phenomenon
of dipole blockade \cite{lufl+:01} which prevents simultaneous
photo-excitation of nearby atoms. On the other hand the large dipole moments
associated with electronic transitions among Rydberg states lead to fast
resonant energy transfer over long distances.

Dipole blockade has been widely exploited for applications in quantum
information processing \cite{sawa+:10} and quantum optics
\cite{prma+:10,sehe+:11,peot+:11,duku:12,pefi+:12,gona+:13}. In the
many-body context, spatial correlations of laser-excited atoms were recently
investigated theoretically in one-dimensional finite-size samples
\cite{atle:12,brsc+:12,gahe+:12,homu+:13}. It was found that at high atomic
density the distribution of the Rydberg atoms can become highly structured
leading in the extreme case to the spontaneous formation of small ordered
patches consisting of a few Rydberg atoms. From the experimental side much
effort has been invested to probe these spatial correlations
\cite{scsa+:11,duba+:12,scch+:12}. In parallel, resonant energy transfer has
been the focus of pioneering experiments on cold (frozen) Rydberg gases
\cite{moco+:98,anve+:98}. Spectral broadening of Rydberg lines could be
attributed to coherent electronic energy transfer between Rydberg atoms.
These findings have ushered further experimental
\cite{lita+:05,diko+:08,niba+:12,masz+:13} as well as theoretical
\cite{frce+:99,rohe+:04,mubl+:07,muli+:08,wuat+:10,wuat+:11,scwe+:11,kipa+:12,badu+:12}
efforts to analyze excitonic transfer processes and their consequences in
cold Rydberg systems.

\begin{figure}
\centering
\includegraphics[width=0.95\columnwidth]{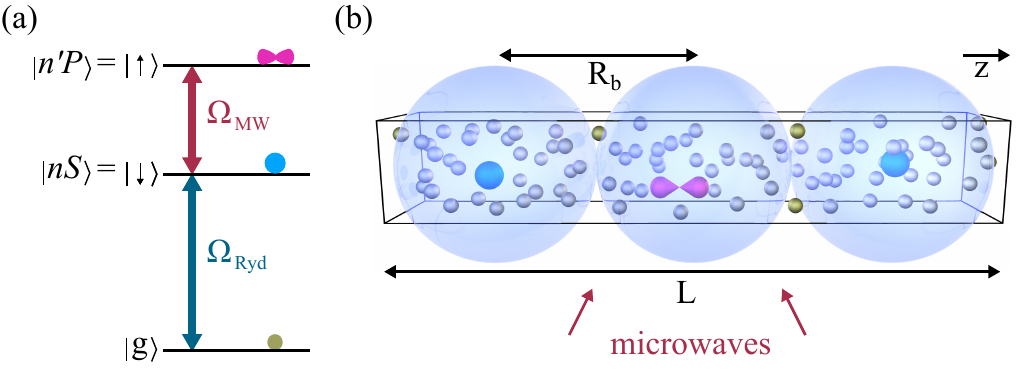}
\caption{(Color online) (a) Rydberg levels and excitation scheme. In step 1
  atoms are optically excited from their electronic ground state $|g\rangle$
  to a highly-excited state $| nS \rangle$ with effective Rabi frequency
  $\Omega_{\text{Ryd}}$. Subsequently, in step 2, a microwave field is
  applied that couples the $|nS\rangle$ state to a nearby $| n^{\prime} P
  \rangle$ level with Rabi frequency $\Omega_{\text{MW}}$. These states are
  identified as internal states of a fictitious spin-1/2 particle. (b) Quasi
  one-dimensional geometry of the system. The large van der Waals potential
  between the $|nS\rangle$ Rydberg atoms (blue spheres) induces an exclusion
  volume around each excited atom characterized by the blockade radius
  $R_{\rm b}$ larger than the transverse size of the sample. This leads to a
  highly structured density distribution of the Rydberg atoms after the
  laser excitation step. The transition dipole-dipole interaction induces a
  coherent spatial transfer of $(n^{\prime}P)$-excitations.}
\label{fig:setup}
\end{figure}

In this work we build on these developments and present a study of the
out-of-equilibrium behavior of Rydberg gases that directly links to the
above-mentioned themes of order formation and excitation transfer. We
explore the electronic dynamics of a (quasi) one-dimensional Rydberg gas
within a two-step protocol, see Fig.\ \ref{fig:setup}. In step 1 high-lying
Rydberg $nS$-states are optically excited leading to the spontaneous
emergence of a lattice of Rydberg atoms immersed in the atomic gas of ground
state atoms. In the subsequent step 2 we trigger coherent excitation
transfer between Rydberg atoms by the excitation of nearby Rydberg $n^\prime
P$-states via a microwave field. We theoretically investigate the resulting
non-equilibrium exciton dynamics through the analysis of the survival
probability of atoms in the Rydberg $nS$-state. We show that the survival
probability has a characteristic dependence on the spatial arrangement as
well as on the number of Rydberg atoms and derive an effective Hamiltonian
in the limit of strong microwave driving. We compare our theoretical
predictions with an experiment in which the survival probability is measured
through an optical read-out protocol and find qualitative agreement. Our
work shows the versatility of homogeneous gases of Rydberg atoms for the
study of non-equilibrium processes such as coherent transport phenomena.

The paper is organized as follows: In section \ref{sec:lattice} we
theoretically describe and investigate the spontaneous formation of lattices
of Rydberg atoms upon photo-excitation from a one-dimensional finite-size
atomic gas. Subsequently, we discuss the electronic dynamics when the
excited atoms are coupled to a nearby Rydberg state by a microwave (section
\ref{sec:excitons}). In particular, we illuminate on the connection between
the electronic dynamics with the underlying ordering of the Rydberg
atoms. Finally, in section \ref{sec:experiment} we present experimental data
that are obtained using a similar protocol as the one discussed in this work
and find qualitative agreement between experiment and theory. Unless stated
otherwise we will work in units where $\hbar=1$.

\section{Emergent lattice}\label{sec:lattice}
First we introduce the framework within which we describe exciton dynamics
in an emergent lattice (cf.\ Fig.\ \ref{fig:setup}). We consider a
one-dimensional atomic gas of length $L$ with homogeneous density
$\rho=N/L$, where $N$ denotes the total number of atoms and we set our
quantization axis ($z$) along the long axis of the gas. In order to describe
the electronic dynamics, we use a simplified level scheme in which each atom
is modeled by using three levels: the electronic ground state $|g\rangle$
and two dipole-coupled, highly excited states $| nS \rangle$ and
$|n^{\prime} P \rangle$ with principal quantum numbers $n$ and $n^{\prime}$
and angular momentum $l=0$ ($S$) and $l=1$ ($P$), respectively. The actual
states used in the experiment will be discussed later.

In the first step of our protocol, atoms are optically excited from
$|g\rangle$ to $|n S\rangle$. The dynamics in this step are dominated by the
strong van der Waals interaction between Rydberg atoms.  Such interactions
shift many-body states containing pairs of Rydberg atoms closer than a
critical distance out of resonance. Consequently each excited atom is
surrounded by a blockade volume of radius $R_b$, within which no further
Rydberg excitations are found. For our finite, one-dimensional, homogeneous
system this dipole blockade mechanism restricts the maximum number of
Rydberg excitations to $\nu_{\text{max}} = \lfloor L/R_b \rfloor+1$, where
$\lfloor x \rfloor$ denotes the closest integer smaller than $x$. We have
assumed a sharply defined blockade volume which is justified in case of a
one-dimensional system and rapidly decaying van der Waals interaction
\cite{atle:12,peho+:13}. The quantum state after the laser excitation (i.e.,
the initial state of the microwave driving) can be formally written as
\begin{equation}
| \Psi_0 \rangle = \sum_{\nu=0}^{\nu_{\text{max}}} 
\sum_{1\le \alpha_1 < \dots < \alpha_{\nu} \le N} g^{(\nu)}_{\alpha_1 \dots \alpha_{\nu}} \left(
\prod_{j=1}^{\nu} e^{-ikz_{\alpha_j}} S^+_{\alpha_j}
\right) |0 \rangle . \label{eq:state}
\end{equation}
Here, $|0\rangle=\bigotimes_i |g\rangle_i$ is the Rydberg vacuum, $S^+_i =
|nS\rangle\langle g|_i$ is the operator that creates a Rydberg atom at
position $z_i$ and $k$ is the total momentum imposed by the excitation
laser(s), where we assume $\hat{k} \parallel \hat{z}$. In general, this
state is highly correlated and the coefficients $g^{(\nu)}_{\alpha_1 \dots
  \alpha_{\nu}} \equiv g^{(\nu)}(z_{\alpha_1}, \dots , z_{\alpha_{\nu}})$
depend on the positions of all excited atoms and on time (the time label was
suppressed to shorten the notation). These correlations, which dynamically
build up during the laser-excitation, render a full quantum calculation of
$|\Psi_0\rangle$ a formidable task.

However, as recently shown numerically \cite{leol+:10} and analytically
\cite{atga+:12,jiat+:13}, for sufficiently large times the strong
interactions between Rydberg atoms result in an equilibration. In this
saturated regime the moduli of the coefficients $g^{(\nu)}$ of
configurations compatible with the Rydberg blockade \emph{become equal and
  constant in time} \cite{atle:12}. Therefore, observables that depend only
on the $|g^{(\nu)}|^2\!\!$, such as the Rydberg density and density-density
correlations, attain a stationary value. To calculate such 'classical
observables' we use a Monte Carlo method that allows us to sample arrangements 
of Rydberg atoms compatible with the excitation blockade. 
The details of the algorithm are given in Appendix \ref{app:algorithm}. 

\begin{figure*}
\includegraphics[width=\textwidth]{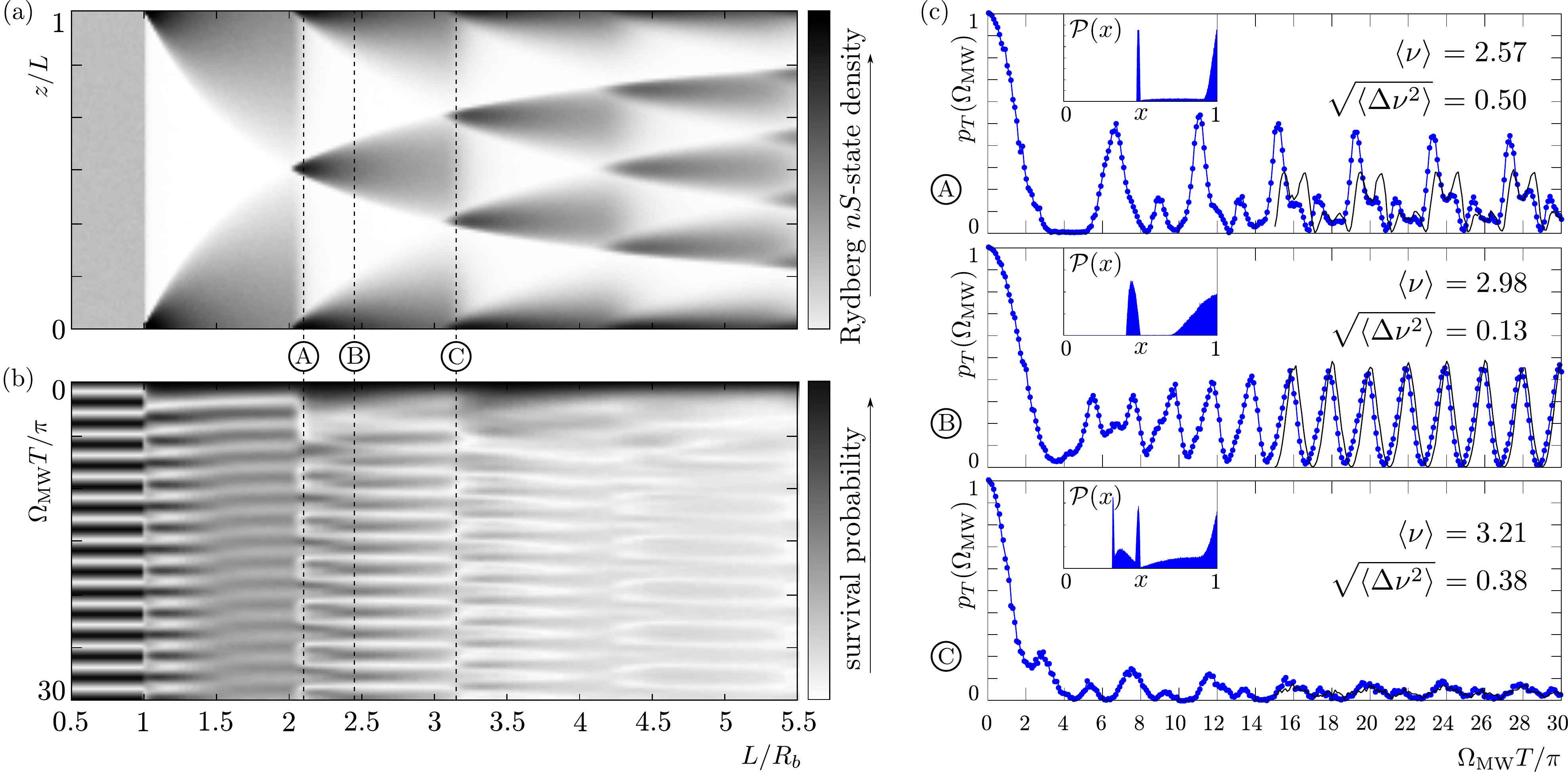}
\caption{(Color online) (a) Numerically calculated density distribution of
  Rydberg atoms after laser excitation as a function of the system length
  $L$ measured in units of the blockade radius $R_b$. (b) Survival
  probability $p_T(\Omega_\text{MW})$ as a function of $L/R_b$ and microwave
  Rabi frequency $\Omega_{\text{MW}}$. The colormaps in panels (a) and (b)
  range from 0 (white) to 1 (black). (c) Cuts through the density plot shown
  in (b) for three different system sizes indicated by the encircled capital
  letters. The corresponding mean, $\langle\nu\rangle$, and variance,
  $\sqrt{\left(\Delta \nu^2\right) }$, of the Rydberg atom number
  distribution are indicated in the panels. The blue symbols show numerical
  results from the Hamiltonian given in eq. (\ref{eq:model_hamil}) and the
  solid black lines are obtained using the effective Hamiltonian
  (\ref{eq:xxz}) valid for large $\Omega_{\text{MW}}$. The statistical
  distribution of the dipole-dipole interaction energy between atom pairs is
  shown in the insets. For better visibility we show the distributions as a
  function of $x=(\mathcal{V} R^3_b/\mu^2)^{1/3}$, i.e. the third root of
  the interaction strength at the blockade distance $R_b$. The data shown in
  panels (b) and (c) were obtained by averaging over 400 initial
  configurations generated by Monte Carlo sampling. In (b) and (c) the
  dipole-dipole interaction strength, $V(R_b)$, for an atom pair separated
  by $R_b$ is $V(R_b)T = 15$.}
\label{fig:results}
\end{figure*}

Fig. \ref{fig:results}(a) shows the numerically calculated stationary
Rydberg density distributions as a function of the system length $L$ for a
gas consisting of $N=10^4$ atoms at fixed blockade radius and open boundary
conditions. For $L>R_b$ the plot shows a highly structured density
distribution with pronounced peaks at the boundaries of the gas. Whenever
the system size slightly exceeds an integer multiple of the blockade radius
the density distribution closely resembles a lattice of Rydberg atoms, i.e.,
the excited atoms are arranged in the densest packing allowed by the
excitation blockade. Similar structures have also been reported in
\cite{gahe+:12,homu+:13,peho+:13}. We emphasize however that the emergence
of such seemingly ordered structures is a finite-size effect, as for
one-dimensional systems with finite-range interactions spatial correlations
decay exponentially with increasing distance in the thermodynamic limit
\cite{atle:12}.

\section{Excitonic dynamics}\label{sec:excitons}
Once the Rydberg lattice is created we trigger the excitonic dynamics by
irradiating the ensemble with a microwave field (linearly polarized along
the $z$-axis and with Rabi frequency $\Omega_\text{MW}$) that is resonant
with the $| n S \rangle$-$| n^{\prime} P \rangle$ transition as shown in
Fig. \ref{fig:setup}(a). The corresponding transition dipole moment, $\mu$,
can reach thousands of atomic units (scaling as $n^2$ for $n=n^{\prime}$).
This results in a strong microwave coupling but also induces a significant
resonant dipole-dipole interaction enabling excitations to be exchanged
coherently between atoms, i.e., two distant Rydberg atoms swap their
electronic state according to $|nS\rangle_i|n^{\prime} P\rangle_j
\leftrightarrow |n^{\prime} P\rangle_i|nS\rangle_j$. For the particular
geometry of our system this resonant dipole-dipole interaction does not
couple different magnetic sub-levels and the electronic structure of each
Rydberg atom reduces to that of an effective spin 1/2 particle where we
identify $| nS \rangle \equiv \left|\downarrow\right>$ and $| n^{\prime} P
\rangle \equiv \left|\uparrow\right>$. In atomic units the interaction
strength is given by $V_{ij} = (-2/3)\mu^2/|z_i-z_j|^3$, where $z_i$ denotes
the position of the $i$-th Rydberg atom \cite{rohe+:04}. The Hamiltonian
describing the resulting dynamics is given by
\begin{equation}
H = \frac{\Omega_{\text{MW}}}{2} \sum_i \sigma_i^x + \sum_{ij, (i\ne j)} V_{ij} 
\left( \sigma_i^+ \sigma_j^- + \sigma_i^- \sigma_j^+ \right)
\label{eq:model_hamil}
\end{equation}
with $\sigma^{\pm} = (\sigma^x \pm {\rm i} \sigma^y)/2$ and $\sigma^{x,y,z}$
denoting the usual Pauli spin matrices.

In order to study the excitonic dynamics we analyze the survival
probability, $p_{T}(\Omega_{\text{MW}}) \equiv |\langle \Psi_0 | e^{-{\rm i}
  HT} |\Psi_0 \rangle |^2$, of the state $|\Psi_0 \rangle$. The motivation
behind choosing this quantity stems from the fact that $p_T(\Omega
_{\text{MW}})$ is accessible experimentally, as has been shown recently
\cite{masz+:13}, and that its rich dynamical behavior reveals interesting
insights into the physics of the driven exciton system. Following
Ref.\ \cite{masz+:13} we numerically determine $p_T(\Omega_{\text{MW}})$ for
varying strength of the microwave coupling and in addition as a function of
the system length for fixed time $T$ and blockade radius $R_{\rm b}$. The
results are summarized in Fig.\ \ref{fig:results}(b), where the
dipole-dipole interaction strength is chosen such that for two atoms
separated by the blockade radius $V(R_b) T =15.$ Additionally, we present
cuts through these data for three selected values of the system length in
Fig.\ \ref{fig:results}(c). The survival probability exhibits a rich
structure as $L$ and $\Omega_{\text{MW}}$ are varied which, in particular,
is clearly correlated to the distribution of Rydberg atoms shown in panel
(a). This is most apparent when the system length is close to an integer
multiple of the blockade radius, i.e., when an additional Rydberg atom can
``fit'' into the system. Here, the change in the Rydberg density
distribution becomes strikingly visible through a phase jump in the survival
probability pattern.

Beyond this global feature, each cut of $p_T(\Omega_{\text{MW}})$, taken at
a fixed system length $L$, exhibits three distinct regimes as a function of
the microwave Rabi frequency [cf.\ panels of Fig.\ \ref{fig:results}(c)].
For very small $\Omega_{\text{MW}}$ the dynamics is dominated by excitonic
energy transfer induced by the dipole-dipole interaction. The system can
then be regarded as a number of exciton bands that are weakly coupled by the
microwave field. Here, $p_T(\Omega_{\text{MW}})$ decreases monotonically
with increasing $\Omega_{\text{MW}}$. When the microwave Rabi frequency
becomes comparable to the mean dipole-dipole energy, $\langle \mathcal{V}
\rangle$, the survival probability exhibits a rather intricate pattern.
Here, as well as in the previous regime the exact details of
$p_T(\Omega_{\text{MW}})$ are very sensitive to the particular distribution
of the Rydberg atoms [cf.\ Fig.\ \ref{fig:results} (a) and (b)]. This
sensitivity is caused by the distance-dependence of the dipole-dipole
interaction, which essentially probes the pair distribution function
$\mathcal{R}^{(2)} (z,z^{\prime})$ of the Rydberg gas. To see this one can
use the formal expression (\ref{eq:state}) for $|\Psi_0\rangle$ to express
the survival probability. The resulting expression depends on the
distributions $\rho^{(\nu)} (z_1,\dots ,z_{\nu}) = |g^{(\nu)} (z_1, \dots ,
z_{\nu})|^2$ in each particle number subspace, which together determine the
pair distribution function through
\begin{equation}
\mathcal{R}^{(2)} (z,z^{\prime}) 
=\sum_{\nu=2}^{\nu_{\text{max}}} \int \text{d} z_3 
\cdots \text{d}z_{\nu} \rho^{(\nu)} (z,z^{\prime},z_3,\dots ,z_{\nu}).
\end{equation}
Since the interaction potential has a strong nonlinear dependence on the
inter-particle distance even small differences in
$\mathcal{R}^{(2)}(z,z^{\prime})$ lead to a significantly modified
statistical distribution
\begin{equation}
\mathcal{P}(\mathcal{V}) = \int \text{d}z 
\text{d}z^{\prime} \delta\left( \mathcal{V} 
- \mu^2/|z-z^{\prime}|^3 \right) \mathcal{R}^{(2)} (z,z^{\prime})
\end{equation}
of the dipole-dipole interaction energy $\mathcal{V}$ as shown in the insets
of the panels in Fig.\ \ref{fig:results}(c).

Finally, for $\Omega_{\text{MW}} \gg \langle \mathcal{V} \rangle$ the
survival probability shows regular oscillations. Most interestingly, as a
general trend the amplitude of these oscillations decreases with increasing
number of Rydberg atoms in the gas [cf.\ rightmost part of
  Fig.\ \ref{fig:results}(b)].  This effect seems counterintuitive as for
$\Omega_{\text{MW}}/\langle \mathcal{V} \rangle \gg 1$ one might expect to
enter a non-interacting regime in which the microwave drives coherent
oscillations between the single atom states $|nS\rangle$ and $|n^{\prime}
P\rangle$. This would always lead to oscillation of $p_T(\Omega
_{\text{MW}})$ with full contrast. However this is a misconception as even
for $\Omega_{\text{MW}}/\langle \mathcal{V} \rangle \gg 1$ interactions
still play a significant role. This can be understood by deriving the
effective Hamiltonian in the limit $V_{ij}\ll\Omega_\mathrm{MW}$, starting
from eq. (\ref{eq:model_hamil}). To this end, we rotate the spin basis via a
unitary transformation $U = \bigotimes_{j} U_j$ with $U_j = \exp({\rm i} \pi
\sigma^y_j/4)$, which diagonalizes the single-body part of $H$. In the
transformed Hamiltonian $UHU^{\dagger}$ we neglect non-resonant terms of the
form $\sigma^+_i \sigma^+_j$ and $\sigma^{-}_i \sigma^{-}_j$ that correspond
to the simultaneous (de-)excitation of a spin pair. Within this
approximation we find that the effective Hamiltonian is that of the
spin-$1/2$ XXZ-model,
\begin{equation}
H_{\text{eff}} =  \frac{\Omega_{\text{MW}}}{2} 
\sum_i \sigma_i^z + \sum_{ij, (i\ne j)} \frac{V_{ij}}{4} \left(
\sigma_i^x \sigma_j^x + \sigma_i^y \sigma_j^y + 2 \sigma_i^z \sigma_j^z
\right).
\label{eq:xxz}
\end{equation}
This Hamiltonian consists of two commuting parts and thus the microwave
driving and the residual dipole-dipole interaction can be treated
independently. This shows that no matter how strong $\Omega_{\text{MW}}$
there will always be a nontrivial excitonic dynamics. To illustrate this
further let us assume that we are in a regime where the number of Rydberg
atoms $\nu_\text{Ryd}$ is integer and hence not fluctuating, as e.g. shown
in the middle panel of Fig.\ \ref{fig:results}(c). The survival amplitude
can now be expressed in terms of a Fourier series, $ \langle \Psi_0 | e^{-i
  H_{\text{eff}} T} | \Psi_0\rangle = B(T) + 2\sum_{m=0}^{M} A_{m} (T)\cos(
\epsilon_{m} T)$, with $\epsilon_{m} = \Omega_{\text{MW}} (m -
\nu_{\text{Ryd}}/2)$ and coefficients $A_{m} (T)$ and $B(T)$ which
exclusively depend on the phases $V_{ij}T$ but not on $\Omega_{\text{MW}}$.
If $\nu_{\text{Ryd}}$ is odd the summation runs to $M=(\nu_{\text{Ryd}}
-1)/2$ otherwise to $M=\nu_{\text{Ryd}}/2 -1$.  In case of odd
$\nu_{\text{Ryd}}$ the symmetry of the Hamiltonian (\ref{eq:xxz}) with
respect to a global spin flip operation imposes $B(T) =0$. Hence, in this
case the survival probability is exactly zero when $\Omega_{\text{MW}} T
/\pi$ is an odd integer, which is clearly visible in the middle panel of
Fig.\ \ref{fig:results}(c). Note, that in general only if $V_{ij}=0$ the
Fourier series can be summed analytically and one consequently obtains the
familiar expression for Rabi oscillations of non-interacting particles
$|\langle \Psi_0 | e^{-i H_{\text{eff}} T} | \Psi_0\rangle|^2
=\cos^{2\nu_{\text{Ryd}}}(\Omega_\text{MW}T)$.

\section{Experiment}\label{sec:experiment}
\begin{figure}
\raggedright
\includegraphics[width=\columnwidth]{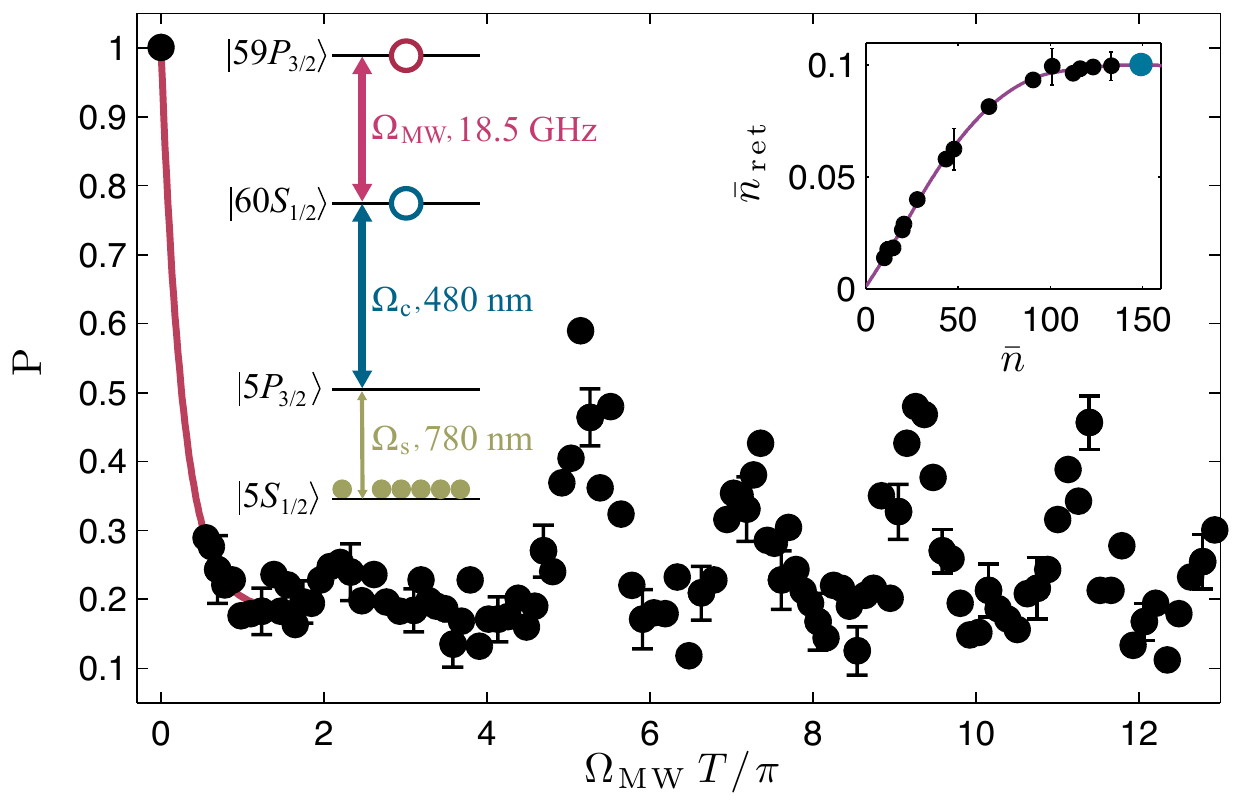}
\caption{(Color online). Experimentally measured photon retrieval
  probability, $\mathrm{P}$, as a function of the microwave Rabi frequency
  $\Omega_\text{MW}$. The duration of the microwave pulse, $T=150$ ns, is
  fixed. Inset: (Left) Excitation scheme used in the experiment. The Rydberg
  $60S_{1/2}$-state is reached by a two-photon transition via the
  $5P_{3/2}$-state. A microwave field couples the $60S_{1/2}$-state to the
  $59P_{3/2}$-state. (Right) As the number of photons $\bar{n}$ in the
  signal field (proportional to the Rabi frequency $\Omega_\mathrm{s}$ and
  pulse duration) is increased, the number of retrieved photons
  $\bar{n}_{\mathrm{ret}}$ saturates in the absence of microwave driving.
  The data shown in the main plot is obtained well within this saturated
  regime (blue point). The lines are a guide to the eye.}
\label{fig:experiment}
\end{figure}

We will now conclude with a qualitative comparison of our theoretical
results with experimental data. Full details of the experiment can be found
in reference \cite{masz+:13}. In brief, laser-cooled $^{87}\mathrm{Rb}$
atoms are loaded into an optical dipole trap forming an elongated cloud with
$w_{z}\approx 20~\mu$m and $w_{\rm{r}}\approx 0.8~\mu$m, where $w$ is the
standard deviation of the density distribution.  Rydberg $60S_{1/2}$-states
are excited with a two-photon transition via the intermediate
$5P_{3/2}$-state (see Fig. \ref{fig:experiment}). For the experimental
parameters the blockade radius is $R_{\rm b}\approx7~\mu\text{m} \gg
w_{\rm{r}}$ and hence the setup is quasi-one-dimensional as illustrated in
Fig. \ref{fig:setup}(b). The optical excitation fields are applied for a
sufficiently long time that saturation is reached (see right inset of
Fig. \ref{fig:experiment} and corresponding caption). This corresponds to
step 1, i.e. the preparation of the emergent lattice. Subsequently, the
lasers are switched off and a microwave field (step 2, driving the exciton
dynamics), couples the states $\left|60S_{1/2}\right>$ and $\vert59P
_{3/2}\rangle$, for a time $T= 150$ ns. After step 2 the Rydberg excitations
in state $60S_{1/2}$ are converted into photons by switching on the control
field, see \cite{masz+:13}. The resulting photon-retrieval probability,
$\mathrm{P}$, is proportional to the survival probability
$p_T(\Omega_{\text{MW}})$ \cite{masz+:13} and plotted in
Fig.~\ref{fig:experiment} as a function of the microwave Rabi frequency,
$\Omega_{\text{MW}}$.

Remarkably, this data shows qualitatively the same features as the discussed
theoretical model, i.e., a quick decay of the survival probability, followed
by increasingly regular oscillations. In fact the shape of the curve follows
quite closely the behavior depicted in the middle panel of
Fig. \ref{fig:results}c. This is despite the fact that details of the
experiment, e.g. the excitation scheme and the Gaussian atomic density
distribution, differ from the underlying theoretical model.

\section{Conclusions}\label{sec:conclusions}
In this work we have studied the dynamics of a two-step protocol that
consists (i) of photo-exciting Rydberg atoms from an one-dimensional
ultracold gas and (ii) subsequently inducing excitonic energy transfer in
the highly correlated Rydberg gas using a microwave field. In the regime of
weak microwave driving we have shown that the exciton dynamics strongly
depends on the spatial arrangement of the Rydberg atoms. This connection
opens up the possibility of using the second step of our protocol as a
diagnostic tool for mapping out the spatial structure of highly correlated
Rydberg gases. In the opposite limit of strong microwave driving we have
shown that the dynamics of the system is governed by the Hamiltonian of the
XXZ model. This highlights the applicability of simple models to describe
and understand complex experiments to investigate non-equilibrium exciton
dynamics in ultracold Rydberg systems.

\begin{acknowledgments}
We thank B.\ Olmos for useful comments on the manuscript. S.B. and
T.F. acknowledge funding from EPSRC. D.M. and C.S.A. acknowledge support
from EPSRC, Durham University and the EU Marie Curie ITN COHERENCE.
I.L. acknowledges funding by EPSRC, the Leverhulme Trust and the EU-FET grant QuILMI 295293. C.A. acknowledges support through the Alexander von Humboldt Foundation.
\end{acknowledgments}

\appendix
\section{Configuration-selection algorithm}
\label{app:algorithm}

The Monte-Carlo algorithm that we have used to obtain our numerical data relies on the generation of excitation configurations (that are compatible with the Rydberg blockade) 'on the fly'. It is composed of a preparation stage and an actual drawing stage. 

\emph{Preparation stage --}
In the first stage we prepare two tables ($c(j,k)$ and $n(j)$) that we will use to quickly generate random configurations in the second stage of our algorithm. To this end we label the $N$ atoms from left to right with an index $j$ and define $c(j,k)$ to be the number of allowed configurations (compatible with the blockade) with $k \geq 0$ excitations in the index range $[j,N]$ and zero excitations in the range $[1,j-1]$. Furthermore, we define $n(j) > j$ as the index of the first atom to the right of a Rydberg atom at position $j$ that lies outside the blockade radius. With these definitions the following recursive relation holds for
$k>1$,
\begin{equation}
  \label{eq:recursive}
  c(j,k) = {\textstyle \sum_{q=j}^N} \,\, c(n(q),k-1),
\end{equation}
with initial conditions $c(j,0)=1$ and \mbox{$c(j,1)=N-j+1$}. This permits
the calculation of the total number $\mathcal{N}$ of allowed configurations
as well as the fraction $p_k$ of configurations with exactly $k$
excitations, according to the formulae $\mathcal{N} = \sum_{k\geq 0} c(1,k)$
and $p_k = c(1,k) / \mathcal{N}$. 

In order to illustrate the procedure, let us take $N=7$ atoms spaced as in the following diagram (where the distance between two vertical bars is one blockade radius)\bigskip

\newcommand{\mybar}{\rlap{\rule{1pt}{2ex}}\hspace{10ex}}
\hspace{1em}\rlap{\raisebox{.5ex}{\rule{.8\linewidth}{1pt}}}%
\rlap{\hspace{2.5ex}\mybar\mybar\mybar\mybar\mybar}%
\hspace{2ex}\rlap{$\bullet$}%
\hspace{6ex}\rlap{$\bullet$}%
\hspace{7ex}\rlap{$\bullet$}%
\hspace{5ex}\rlap{$\bullet$}%
\hspace{11ex}\rlap{$\bullet$}%
\hspace{5ex}\rlap{$\bullet$}%
\hspace{6ex}\rlap{$\bullet$}%
\\\hspace*{1em}\hspace{2ex}\hspace{.3ex}%
\hspace{2ex}\rlap{1}
\hspace{6ex}\rlap{2}
\hspace{7ex}\rlap{3}
\hspace{5ex}\rlap{4}
\hspace{11ex}\rlap{5}
\hspace{5ex}\rlap{6}
\hspace{6ex}\rlap{7}
\bigskip

The corresponding values of $n(j)$ and $c(j,k)$ are then
\begin{center}
  \hfill
  \begin{tabular}{l|c}
    $j$~~ & $n(j)$ \\ \hline 1&3 \\[-.5ex] 2&4 \\[-.5ex] 3&5 
    \\[-.5ex] 4&5 \\[-.5ex] 5&7 \\[-.5ex] 6& \\[-.5ex] 7& \\ ~
  \end{tabular}
  \hfill
  \begin{tabular}{r|@{\hspace{2ex}}*{5}{c}}
    $j\backslash^{\!\displaystyle k}$ 
        & 0 & 1 &  2 &  3 & 4 \\ \hline
    1~~ & 1 & 7 & 16 & 13 & 3 \\[-.5ex]
    2~~ & 1 & 6 & 11 &  6 & 1 \\[-.5ex]
    3~~ & 1 & 5 &  7 &  2     \\[-.5ex]
    4~~ & 1 & 4 &  4 &  1     \\[-.5ex]
    5~~ & 1 & 3 &  1          \\[-.5ex]
    6~~ & 1 & 2               \\[-.5ex]
    7~~ & 1 & 1               \\
    $p_k$ & $\tfrac{1}{40}$ & $\tfrac{7}{40}$ & $\tfrac{16}{40}$ 
    & $\tfrac{13}{40}$ & $\tfrac{3}{40}$
  \end{tabular}
  \hfill~
\end{center}
Therefore, $\nu_{\text{max}} = 4$, $\mathcal{N}=\sum_k c(1,k) = 40$. 

Note that the $\mathcal{N}$ actual configurations are never explicitly generated (as the
resources in time and memory would scale exponentially with $N$), but for
sake of clarity the $40$ possible ones are:
\begin{center} \small
\begin{tabular}{c@{~\vline~}c@{~\vline~}l}
  \# & \# & actual configurations \\ excit & config & 
  (excited atoms shown in parentheses) \\[1ex] \hline
  0 & 1  & (); \\
  1 & 7  & (1), (2), (3), (4), (5), (6), (7); \\
  2 & 16 & (13), (14), (15), (16), (17), (24), (25), (26), \\
    &    & (27), (35), (36), (37), (45), (46), (47), (57); \\
  3 & 13 & (135),(136),(137),(145),(146),(147),(157), \\
    &    & (245),(246),(247),(257),(357),(457); \\
  4 & 3  & (1357), (1457), (2457).
\end{tabular}
\end{center}

\emph{Drawing stage --}
Once the $c(j,k)$ table is ready, we can use it to randomly sample the space of allowed configurations. To this end we draw a random integer $m$ in the range $[1,\mathcal{N}]$.  In order to uniquely ascribe a specific configuration to $m$ we use the convention that configurations are first ordered according to
the number $\nu$ of excitations and then lexicographically over the indexes of excited atoms. The search is implemented first scanning the entries $c(1,k)$ till $\nu$ is determined, then using the entries $c(j,\nu)$ to find the actual excitation positions. In \texttt{C} pseudo-code the algorithm reads:

\noindent%
\begin{verbatim}
k=0; j=1;
while (m>c(1,k)) {m-=c(1,k); ++k;} // step 1
while (k>1) { q=j; // step 2
  while (m>c(n(q),k-1)) {m-=c(n(q),k-1); ++q;}
  mark_as_excited(q); j=n(q); --k; }
if (k>0) mark_as_excited(j+m-1); 
\end{verbatim}

Let us see how this works in the example for $m=35$:
\begin{itemize}
\setlength{\itemsep}{-\parskip}
\newcommand{\showm}[1]{{\small[$m$=\makebox[2.5ex]{#1}]}}
\item[\showm{35}] $m>c(1,0)=1 \Rightarrow$ not the configuration with zero excitations, then update $m \rightarrow m-c(1,0)=34$; 
\item[\showm{34}] $m>c(1,1)=7 \Rightarrow$ not a configuration with one excitation, then update $m \rightarrow m-c(1,1)=27$; 
\item[\showm{27}] $m>c(1,2)=16 \Rightarrow$ not a configuration with two excitations, then update $m \rightarrow m-c(1,2)=11$; 
\item[\showm{11}] $m\leq c(1,3)=13 \Rightarrow$ the selected configuration has $\nu = 3$ excitations; let us look for their indexes;
\item[\showm{11}] $m>c(n(1),\nu-1)=7 \Rightarrow$ the first index is larger than 1, then update $m \rightarrow m-c(n(1),\nu-1)=4$;
\item[\showm{4}] $m\leq c(n(2),\nu-1)=4 \Rightarrow$ the first index is $j_1=2$; let us look for the second one (from $n(j_1)=4$ on);
\item[\showm{4}] $m>c(n(4),\nu-2)=3 \Rightarrow$ the second index is larger than 4, then update $m \rightarrow m-c(n(4),\nu-2)=1$;
\item[\showm{1}] $m\leq c(n(5),\nu-2)=1 \Rightarrow$ the second index is $j_2=5$; let us look for the last one;
\item[\showm{1}] the last index is just $n(j_2)+m-1=7+1-1=7$, then the selected configuration is (257). 
\end{itemize}

Using this procedure repeatedly, we can generate a random sequence $\mathcal{S} = \{ m_1,m_2,\dots , m_K\}$ of $K$ allowed configuration, where each configuration is drawn with equal probability $1/\mathcal{N}$. Classical observables like the local Rydberg density $\langle n_i \rangle$ or density-density correlations $\langle n_i n_j \rangle$ can then be determined as
\begin{eqnarray}
\langle n_i \rangle &\approx & \frac{1}{K} \sum_{\mathcal{S}} \left( n_i \right)_{\mathcal{S}} \\
\langle n_i n_j \rangle &\approx & \frac{1}{K} \sum_{\mathcal{S}} \left( n_i n_j \right)_{\mathcal{S}} .
\end{eqnarray}


\begin{thebibliography}{41}%
\makeatletter
\providecommand \@ifxundefined [1]{%
 \@ifx{#1\undefined}
}%
\providecommand \@ifnum [1]{%
 \ifnum #1\expandafter \@firstoftwo
 \else \expandafter \@secondoftwo
 \fi
}%
\providecommand \@ifx [1]{%
 \ifx #1\expandafter \@firstoftwo
 \else \expandafter \@secondoftwo
 \fi
}%
\providecommand \natexlab [1]{#1}%
\providecommand \enquote  [1]{``#1''}%
\providecommand \bibnamefont  [1]{#1}%
\providecommand \bibfnamefont [1]{#1}%
\providecommand \citenamefont [1]{#1}%
\providecommand \href@noop [0]{\@secondoftwo}%
\providecommand \href [0]{\begingroup \@sanitize@url \@href}%
\providecommand \@href[1]{\@@startlink{#1}\@@href}%
\providecommand \@@href[1]{\endgroup#1\@@endlink}%
\providecommand \@sanitize@url [0]{\catcode `\\12\catcode `\$12\catcode
  `\&12\catcode `\#12\catcode `\^12\catcode `\_12\catcode `\%12\relax}%
\providecommand \@@startlink[1]{}%
\providecommand \@@endlink[0]{}%
\providecommand \url  [0]{\begingroup\@sanitize@url \@url }%
\providecommand \@url [1]{\endgroup\@href {#1}{\urlprefix }}%
\providecommand \urlprefix  [0]{URL }%
\providecommand \Eprint [0]{\href }%
\providecommand \doibase [0]{http://dx.doi.org/}%
\providecommand \selectlanguage [0]{\@gobble}%
\providecommand \bibinfo  [0]{\@secondoftwo}%
\providecommand \bibfield  [0]{\@secondoftwo}%
\providecommand \translation [1]{[#1]}%
\providecommand \BibitemOpen [0]{}%
\providecommand \bibitemStop [0]{}%
\providecommand \bibitemNoStop [0]{.\EOS\space}%
\providecommand \EOS [0]{\spacefactor3000\relax}%
\providecommand \BibitemShut  [1]{\csname bibitem#1\endcsname}%
\let\auto@bib@innerbib\@empty
\bibitem [{\citenamefont {Cross}\ and\ \citenamefont
  {Hohenberg}(1993)}]{crho:93}%
  \BibitemOpen
  \bibfield  {author} {\bibinfo {author} {\bibfnamefont {M.~C.}\ \bibnamefont
  {Cross}}\ and\ \bibinfo {author} {\bibfnamefont {P.~C.}\ \bibnamefont
  {Hohenberg}},\ }\href {\doibase 10.1103/RevModPhys.65.851} {\bibfield
  {journal} {\bibinfo  {journal} {Rev. Mod. Phys.}\ }\textbf {\bibinfo {volume}
  {65}},\ \bibinfo {pages} {851} (\bibinfo {year} {1993})}\BibitemShut
  {NoStop}%
\bibitem [{\citenamefont {Scholes}(2003)}]{sc:03}%
  \BibitemOpen
  \bibfield  {author} {\bibinfo {author} {\bibfnamefont {G.~D.}\ \bibnamefont
  {Scholes}},\ }\href {\doibase 10.1146/annurev.physchem.54.011002.103746}
  {\bibfield  {journal} {\bibinfo  {journal} {Annual Review of Physical
  Chemistry}\ }\textbf {\bibinfo {volume} {54}},\ \bibinfo {pages} {57}
  (\bibinfo {year} {2003})}\BibitemShut {NoStop}%
\bibitem [{\citenamefont {Engel}\ \emph {et~al.}(2007)\citenamefont {Engel},
  \citenamefont {Calhoun}, \citenamefont {Read}, \citenamefont {Ahn},
  \citenamefont {Mancal}, \citenamefont {Cheng}, \citenamefont {Blankenship},\
  and\ \citenamefont {Fleming}}]{enca+:07}%
  \BibitemOpen
  \bibfield  {author} {\bibinfo {author} {\bibfnamefont {G.~S.}\ \bibnamefont
  {Engel}}, \bibinfo {author} {\bibfnamefont {T.~R.}\ \bibnamefont {Calhoun}},
  \bibinfo {author} {\bibfnamefont {E.~L.}\ \bibnamefont {Read}}, \bibinfo
  {author} {\bibfnamefont {T.-K.}\ \bibnamefont {Ahn}}, \bibinfo {author}
  {\bibfnamefont {T.}~\bibnamefont {Mancal}}, \bibinfo {author} {\bibfnamefont
  {Y.-C.}\ \bibnamefont {Cheng}}, \bibinfo {author} {\bibfnamefont {R.~E.}\
  \bibnamefont {Blankenship}}, \ and\ \bibinfo {author} {\bibfnamefont {G.~R.}\
  \bibnamefont {Fleming}},\ }\href {http://dx.doi.org/10.1038/nature05678}
  {\bibfield  {journal} {\bibinfo  {journal} {Nature}\ }\textbf {\bibinfo
  {volume} {446}},\ \bibinfo {pages} {782} (\bibinfo {year}
  {2007})}\BibitemShut {NoStop}%
\bibitem [{\citenamefont {Scholes}\ and\ \citenamefont
  {Rumbles}(2006)}]{scru:06}%
  \BibitemOpen
  \bibfield  {author} {\bibinfo {author} {\bibfnamefont {G.~D.}\ \bibnamefont
  {Scholes}}\ and\ \bibinfo {author} {\bibfnamefont {G.}~\bibnamefont
  {Rumbles}},\ }\href {http://dx.doi.org/10.1038/nmat1710} {\bibfield
  {journal} {\bibinfo  {journal} {Nat Mater}\ }\textbf {\bibinfo {volume}
  {5}},\ \bibinfo {pages} {683} (\bibinfo {year} {2006})}\BibitemShut {NoStop}%
\bibitem [{\citenamefont {Mottl}\ \emph {et~al.}(2012)\citenamefont {Mottl},
  \citenamefont {Brennecke}, \citenamefont {Baumann}, \citenamefont {Landig},
  \citenamefont {Donner},\ and\ \citenamefont {Esslinger}}]{mobr+:12}%
  \BibitemOpen
  \bibfield  {author} {\bibinfo {author} {\bibfnamefont {R.}~\bibnamefont
  {Mottl}}, \bibinfo {author} {\bibfnamefont {F.}~\bibnamefont {Brennecke}},
  \bibinfo {author} {\bibfnamefont {K.}~\bibnamefont {Baumann}}, \bibinfo
  {author} {\bibfnamefont {R.}~\bibnamefont {Landig}}, \bibinfo {author}
  {\bibfnamefont {T.}~\bibnamefont {Donner}}, \ and\ \bibinfo {author}
  {\bibfnamefont {T.}~\bibnamefont {Esslinger}},\ }\href {\doibase
  10.1126/science.1220314} {\bibfield  {journal} {\bibinfo  {journal}
  {Science}\ }\textbf {\bibinfo {volume} {336}},\ \bibinfo {pages} {1570}
  (\bibinfo {year} {2012})}\BibitemShut {NoStop}%
\bibitem [{\citenamefont {Gring}\ \emph {et~al.}(2012)\citenamefont {Gring},
  \citenamefont {Kuhnert}, \citenamefont {Langen}, \citenamefont {Kitagawa},
  \citenamefont {Rauer}, \citenamefont {Schreitl}, \citenamefont {Mazets},
  \citenamefont {Smith}, \citenamefont {Demler},\ and\ \citenamefont
  {Schmiedmayer}}]{grku+:12}%
  \BibitemOpen
  \bibfield  {author} {\bibinfo {author} {\bibfnamefont {M.}~\bibnamefont
  {Gring}}, \bibinfo {author} {\bibfnamefont {M.}~\bibnamefont {Kuhnert}},
  \bibinfo {author} {\bibfnamefont {T.}~\bibnamefont {Langen}}, \bibinfo
  {author} {\bibfnamefont {T.}~\bibnamefont {Kitagawa}}, \bibinfo {author}
  {\bibfnamefont {B.}~\bibnamefont {Rauer}}, \bibinfo {author} {\bibfnamefont
  {M.}~\bibnamefont {Schreitl}}, \bibinfo {author} {\bibfnamefont
  {I.}~\bibnamefont {Mazets}}, \bibinfo {author} {\bibfnamefont {D.~A.}\
  \bibnamefont {Smith}}, \bibinfo {author} {\bibfnamefont {E.}~\bibnamefont
  {Demler}}, \ and\ \bibinfo {author} {\bibfnamefont {J.}~\bibnamefont
  {Schmiedmayer}},\ }\href {\doibase 10.1126/science.1224953} {\bibfield
  {journal} {\bibinfo  {journal} {Science}\ }\textbf {\bibinfo {volume}
  {337}},\ \bibinfo {pages} {1318} (\bibinfo {year} {2012})}\BibitemShut
  {NoStop}%
\bibitem [{\citenamefont {Trotzky}\ \emph {et~al.}(2012)\citenamefont
  {Trotzky}, \citenamefont {Chen}, \citenamefont {Flesch}, \citenamefont
  {McCulloch}, \citenamefont {Schollwock}, \citenamefont {Eisert},\ and\
  \citenamefont {Bloch}}]{trch+:12}%
  \BibitemOpen
  \bibfield  {author} {\bibinfo {author} {\bibfnamefont {S.}~\bibnamefont
  {Trotzky}}, \bibinfo {author} {\bibfnamefont {Y.-A.}\ \bibnamefont {Chen}},
  \bibinfo {author} {\bibfnamefont {A.}~\bibnamefont {Flesch}}, \bibinfo
  {author} {\bibfnamefont {I.~P.}\ \bibnamefont {McCulloch}}, \bibinfo {author}
  {\bibfnamefont {U.}~\bibnamefont {Schollwock}}, \bibinfo {author}
  {\bibfnamefont {J.}~\bibnamefont {Eisert}}, \ and\ \bibinfo {author}
  {\bibfnamefont {I.}~\bibnamefont {Bloch}},\ }\href
  {http://dx.doi.org/10.1038/nphys2232} {\bibfield  {journal} {\bibinfo
  {journal} {Nat Phys}\ }\textbf {\bibinfo {volume} {8}},\ \bibinfo {pages}
  {325} (\bibinfo {year} {2012})}\BibitemShut {NoStop}%
\bibitem [{\citenamefont {Lukin}\ \emph {et~al.}(2001)\citenamefont {Lukin},
  \citenamefont {Fleischhauer}, \citenamefont {C\^ot\'e}, \citenamefont {Duan},
  \citenamefont {Jaksch}, \citenamefont {Cirac},\ and\ \citenamefont
  {Zoller}}]{lufl+:01}%
  \BibitemOpen
  \bibfield  {author} {\bibinfo {author} {\bibfnamefont {M.~D.}\ \bibnamefont
  {Lukin}}, \bibinfo {author} {\bibfnamefont {M.}~\bibnamefont {Fleischhauer}},
  \bibinfo {author} {\bibfnamefont {R.}~\bibnamefont {C\^ot\'e}}, \bibinfo
  {author} {\bibfnamefont {L.~M.}\ \bibnamefont {Duan}}, \bibinfo {author}
  {\bibfnamefont {D.}~\bibnamefont {Jaksch}}, \bibinfo {author} {\bibfnamefont
  {J.~I.}\ \bibnamefont {Cirac}}, \ and\ \bibinfo {author} {\bibfnamefont
  {P.}~\bibnamefont {Zoller}},\ }\href {\doibase 10.1103/PhysRevLett.87.037901}
  {\bibfield  {journal} {\bibinfo  {journal} {Phys. Rev. Lett.}\ }\textbf
  {\bibinfo {volume} {87}},\ \bibinfo {pages} {037901} (\bibinfo {year}
  {2001})}\BibitemShut {NoStop}%
\bibitem [{\citenamefont {Saffman}\ \emph {et~al.}(2010)\citenamefont
  {Saffman}, \citenamefont {Walker},\ and\ \citenamefont
  {M\o{}lmer}}]{sawa+:10}%
  \BibitemOpen
  \bibfield  {author} {\bibinfo {author} {\bibfnamefont {M.}~\bibnamefont
  {Saffman}}, \bibinfo {author} {\bibfnamefont {T.~G.}\ \bibnamefont {Walker}},
  \ and\ \bibinfo {author} {\bibfnamefont {K.}~\bibnamefont {M\o{}lmer}},\
  }\href {\doibase 10.1103/RevModPhys.82.2313} {\bibfield  {journal} {\bibinfo
  {journal} {Rev. Mod. Phys.}\ }\textbf {\bibinfo {volume} {82}},\ \bibinfo
  {pages} {2313} (\bibinfo {year} {2010})}\BibitemShut {NoStop}%
\bibitem [{\citenamefont {Pritchard}\ \emph {et~al.}(2010)\citenamefont
  {Pritchard}, \citenamefont {Maxwell}, \citenamefont {Gauguet}, \citenamefont
  {Weatherill}, \citenamefont {Jones},\ and\ \citenamefont {Adams}}]{prma+:10}%
  \BibitemOpen
  \bibfield  {author} {\bibinfo {author} {\bibfnamefont {J.~D.}\ \bibnamefont
  {Pritchard}}, \bibinfo {author} {\bibfnamefont {D.}~\bibnamefont {Maxwell}},
  \bibinfo {author} {\bibfnamefont {A.}~\bibnamefont {Gauguet}}, \bibinfo
  {author} {\bibfnamefont {K.~J.}\ \bibnamefont {Weatherill}}, \bibinfo
  {author} {\bibfnamefont {M.~P.~A.}\ \bibnamefont {Jones}}, \ and\ \bibinfo
  {author} {\bibfnamefont {C.~S.}\ \bibnamefont {Adams}},\ }\href {\doibase
  10.1103/PhysRevLett.105.193603} {\bibfield  {journal} {\bibinfo  {journal}
  {Phys. Rev. Lett.}\ }\textbf {\bibinfo {volume} {105}},\ \bibinfo {pages}
  {193603} (\bibinfo {year} {2010})}\BibitemShut {NoStop}%
\bibitem [{\citenamefont {Sevin\ifmmode~\mbox{\c{c}}\else \c{c}\fi{}li}\ \emph
  {et~al.}(2011)\citenamefont {Sevin\ifmmode~\mbox{\c{c}}\else \c{c}\fi{}li},
  \citenamefont {Henkel}, \citenamefont {Ates},\ and\ \citenamefont
  {Pohl}}]{sehe+:11}%
  \BibitemOpen
  \bibfield  {author} {\bibinfo {author} {\bibfnamefont {S.}~\bibnamefont
  {Sevin\ifmmode~\mbox{\c{c}}\else \c{c}\fi{}li}}, \bibinfo {author}
  {\bibfnamefont {N.}~\bibnamefont {Henkel}}, \bibinfo {author} {\bibfnamefont
  {C.}~\bibnamefont {Ates}}, \ and\ \bibinfo {author} {\bibfnamefont
  {T.}~\bibnamefont {Pohl}},\ }\href {\doibase 10.1103/PhysRevLett.107.153001}
  {\bibfield  {journal} {\bibinfo  {journal} {Phys. Rev. Lett.}\ }\textbf
  {\bibinfo {volume} {107}},\ \bibinfo {pages} {153001} (\bibinfo {year}
  {2011})}\BibitemShut {NoStop}%
\bibitem [{\citenamefont {Petrosyan}\ \emph {et~al.}(2011)\citenamefont
  {Petrosyan}, \citenamefont {Otterbach},\ and\ \citenamefont
  {Fleischhauer}}]{peot+:11}%
  \BibitemOpen
  \bibfield  {author} {\bibinfo {author} {\bibfnamefont {D.}~\bibnamefont
  {Petrosyan}}, \bibinfo {author} {\bibfnamefont {J.}~\bibnamefont
  {Otterbach}}, \ and\ \bibinfo {author} {\bibfnamefont {M.}~\bibnamefont
  {Fleischhauer}},\ }\href {\doibase 10.1103/PhysRevLett.107.213601} {\bibfield
   {journal} {\bibinfo  {journal} {Phys. Rev. Lett.}\ }\textbf {\bibinfo
  {volume} {107}},\ \bibinfo {pages} {213601} (\bibinfo {year}
  {2011})}\BibitemShut {NoStop}%
\bibitem [{\citenamefont {Dudin}\ and\ \citenamefont
  {Kuzmich}(2012)}]{duku:12}%
  \BibitemOpen
  \bibfield  {author} {\bibinfo {author} {\bibfnamefont {Y.~O.}\ \bibnamefont
  {Dudin}}\ and\ \bibinfo {author} {\bibfnamefont {A.}~\bibnamefont
  {Kuzmich}},\ }\href {\doibase 10.1126/science.1217901} {\bibfield  {journal}
  {\bibinfo  {journal} {Science}\ }\textbf {\bibinfo {volume} {336}},\ \bibinfo
  {pages} {887} (\bibinfo {year} {2012})}\BibitemShut {NoStop}%
\bibitem [{\citenamefont {Peyronel}\ \emph {et~al.}(2012)\citenamefont
  {Peyronel}, \citenamefont {Firstenberg}, \citenamefont {Liang}, \citenamefont
  {Hofferberth}, \citenamefont {Gorshkov}, \citenamefont {Pohl}, \citenamefont
  {Lukin},\ and\ \citenamefont {Vuletic}}]{pefi+:12}%
  \BibitemOpen
  \bibfield  {author} {\bibinfo {author} {\bibfnamefont {T.}~\bibnamefont
  {Peyronel}}, \bibinfo {author} {\bibfnamefont {O.}~\bibnamefont
  {Firstenberg}}, \bibinfo {author} {\bibfnamefont {Q.-Y.}\ \bibnamefont
  {Liang}}, \bibinfo {author} {\bibfnamefont {S.}~\bibnamefont {Hofferberth}},
  \bibinfo {author} {\bibfnamefont {A.~V.}\ \bibnamefont {Gorshkov}}, \bibinfo
  {author} {\bibfnamefont {T.}~\bibnamefont {Pohl}}, \bibinfo {author}
  {\bibfnamefont {M.~D.}\ \bibnamefont {Lukin}}, \ and\ \bibinfo {author}
  {\bibfnamefont {V.}~\bibnamefont {Vuletic}},\ }\href
  {http://dx.doi.org/10.1038/nature11361} {\bibfield  {journal} {\bibinfo
  {journal} {Nature}\ }\textbf {\bibinfo {volume} {488}},\ \bibinfo {pages}
  {57} (\bibinfo {year} {2012})}\BibitemShut {NoStop}%
\bibitem [{\citenamefont {Gorshkov}\ \emph {et~al.}(2013)\citenamefont
  {Gorshkov}, \citenamefont {Nath},\ and\ \citenamefont {Pohl}}]{gona+:13}%
  \BibitemOpen
  \bibfield  {author} {\bibinfo {author} {\bibfnamefont {A.~V.}\ \bibnamefont
  {Gorshkov}}, \bibinfo {author} {\bibfnamefont {R.}~\bibnamefont {Nath}}, \
  and\ \bibinfo {author} {\bibfnamefont {T.}~\bibnamefont {Pohl}},\ }\href
  {\doibase 10.1103/PhysRevLett.110.153601} {\bibfield  {journal} {\bibinfo
  {journal} {Phys. Rev. Lett.}\ }\textbf {\bibinfo {volume} {110}},\ \bibinfo
  {pages} {153601} (\bibinfo {year} {2013})}\BibitemShut {NoStop}%
\bibitem [{\citenamefont {Ates}\ and\ \citenamefont
  {Lesanovsky}(2012)}]{atle:12}%
  \BibitemOpen
  \bibfield  {author} {\bibinfo {author} {\bibfnamefont {C.}~\bibnamefont
  {Ates}}\ and\ \bibinfo {author} {\bibfnamefont {I.}~\bibnamefont
  {Lesanovsky}},\ }\href {\doibase 10.1103/PhysRevA.86.013408} {\bibfield
  {journal} {\bibinfo  {journal} {Phys. Rev. A}\ }\textbf {\bibinfo {volume}
  {86}},\ \bibinfo {pages} {013408} (\bibinfo {year} {2012})}\BibitemShut
  {NoStop}%
\bibitem [{\citenamefont {Breyel}\ \emph {et~al.}(2012)\citenamefont {Breyel},
  \citenamefont {Schmidt},\ and\ \citenamefont {Komnik}}]{brsc+:12}%
  \BibitemOpen
  \bibfield  {author} {\bibinfo {author} {\bibfnamefont {D.}~\bibnamefont
  {Breyel}}, \bibinfo {author} {\bibfnamefont {T.~L.}\ \bibnamefont {Schmidt}},
  \ and\ \bibinfo {author} {\bibfnamefont {A.}~\bibnamefont {Komnik}},\ }\href
  {\doibase 10.1103/PhysRevA.86.023405} {\bibfield  {journal} {\bibinfo
  {journal} {Phys. Rev. A}\ }\textbf {\bibinfo {volume} {86}},\ \bibinfo
  {pages} {023405} (\bibinfo {year} {2012})}\BibitemShut {NoStop}%
\bibitem [{\citenamefont {G\"arttner}\ \emph {et~al.}(2012)\citenamefont
  {G\"arttner}, \citenamefont {Heeg}, \citenamefont {Gasenzer},\ and\
  \citenamefont {Evers}}]{gahe+:12}%
  \BibitemOpen
  \bibfield  {author} {\bibinfo {author} {\bibfnamefont {M.}~\bibnamefont
  {G\"arttner}}, \bibinfo {author} {\bibfnamefont {K.~P.}\ \bibnamefont
  {Heeg}}, \bibinfo {author} {\bibfnamefont {T.}~\bibnamefont {Gasenzer}}, \
  and\ \bibinfo {author} {\bibfnamefont {J.}~\bibnamefont {Evers}},\ }\href
  {\doibase 10.1103/PhysRevA.86.033422} {\bibfield  {journal} {\bibinfo
  {journal} {Phys. Rev. A}\ }\textbf {\bibinfo {volume} {86}},\ \bibinfo
  {pages} {033422} (\bibinfo {year} {2012})}\BibitemShut {NoStop}%
\bibitem [{\citenamefont {H\"oning}\ \emph {et~al.}(2013)\citenamefont
  {H\"oning}, \citenamefont {Muth}, \citenamefont {Petrosyan},\ and\
  \citenamefont {Fleischhauer}}]{homu+:13}%
  \BibitemOpen
  \bibfield  {author} {\bibinfo {author} {\bibfnamefont {M.}~\bibnamefont
  {H\"oning}}, \bibinfo {author} {\bibfnamefont {D.}~\bibnamefont {Muth}},
  \bibinfo {author} {\bibfnamefont {D.}~\bibnamefont {Petrosyan}}, \ and\
  \bibinfo {author} {\bibfnamefont {M.}~\bibnamefont {Fleischhauer}},\ }\href
  {\doibase 10.1103/PhysRevA.87.023401} {\bibfield  {journal} {\bibinfo
  {journal} {Phys. Rev. A}\ }\textbf {\bibinfo {volume} {87}},\ \bibinfo
  {pages} {023401} (\bibinfo {year} {2013})}\BibitemShut {NoStop}%
\bibitem [{\citenamefont {Schwarzkopf}\ \emph {et~al.}(2011)\citenamefont
  {Schwarzkopf}, \citenamefont {Sapiro},\ and\ \citenamefont
  {Raithel}}]{scsa+:11}%
  \BibitemOpen
  \bibfield  {author} {\bibinfo {author} {\bibfnamefont {A.}~\bibnamefont
  {Schwarzkopf}}, \bibinfo {author} {\bibfnamefont {R.~E.}\ \bibnamefont
  {Sapiro}}, \ and\ \bibinfo {author} {\bibfnamefont {G.}~\bibnamefont
  {Raithel}},\ }\href {\doibase 10.1103/PhysRevLett.107.103001} {\bibfield
  {journal} {\bibinfo  {journal} {Phys. Rev. Lett.}\ }\textbf {\bibinfo
  {volume} {107}},\ \bibinfo {pages} {103001} (\bibinfo {year}
  {2011})}\BibitemShut {NoStop}%
\bibitem [{\citenamefont {Dudin}\ \emph {et~al.}(2012)\citenamefont {Dudin},
  \citenamefont {Bariani},\ and\ \citenamefont {Kuzmich}}]{duba+:12}%
  \BibitemOpen
  \bibfield  {author} {\bibinfo {author} {\bibfnamefont {Y.~O.}\ \bibnamefont
  {Dudin}}, \bibinfo {author} {\bibfnamefont {F.}~\bibnamefont {Bariani}}, \
  and\ \bibinfo {author} {\bibfnamefont {A.}~\bibnamefont {Kuzmich}},\ }\href
  {\doibase 10.1103/PhysRevLett.109.133602} {\bibfield  {journal} {\bibinfo
  {journal} {Phys. Rev. Lett.}\ }\textbf {\bibinfo {volume} {109}},\ \bibinfo
  {pages} {133602} (\bibinfo {year} {2012})}\BibitemShut {NoStop}%
\bibitem [{\citenamefont {Schau\ss}\ \emph {et~al.}(2012)\citenamefont
  {Schau\ss}, \citenamefont {Cheneau}, \citenamefont {Endres}, \citenamefont
  {Fukuhara}, \citenamefont {Hild}, \citenamefont {Omran}, \citenamefont
  {Pohl}, \citenamefont {Gross}, \citenamefont {Kuhr},\ and\ \citenamefont
  {Bloch}}]{scch+:12}%
  \BibitemOpen
  \bibfield  {author} {\bibinfo {author} {\bibfnamefont {P.}~\bibnamefont
  {Schau\ss}}, \bibinfo {author} {\bibfnamefont {M.}~\bibnamefont {Cheneau}},
  \bibinfo {author} {\bibfnamefont {M.}~\bibnamefont {Endres}}, \bibinfo
  {author} {\bibfnamefont {T.}~\bibnamefont {Fukuhara}}, \bibinfo {author}
  {\bibfnamefont {S.}~\bibnamefont {Hild}}, \bibinfo {author} {\bibfnamefont
  {A.}~\bibnamefont {Omran}}, \bibinfo {author} {\bibfnamefont
  {T.}~\bibnamefont {Pohl}}, \bibinfo {author} {\bibfnamefont {C.}~\bibnamefont
  {Gross}}, \bibinfo {author} {\bibfnamefont {S.}~\bibnamefont {Kuhr}}, \ and\
  \bibinfo {author} {\bibfnamefont {I.}~\bibnamefont {Bloch}},\ }\href
  {http://dx.doi.org/10.1038/nature11596} {\bibfield  {journal} {\bibinfo
  {journal} {Nature}\ }\textbf {\bibinfo {volume} {491}},\ \bibinfo {pages}
  {87} (\bibinfo {year} {2012})}\BibitemShut {NoStop}%
\bibitem [{\citenamefont {Mourachko}\ \emph {et~al.}(1998)\citenamefont
  {Mourachko}, \citenamefont {Comparat}, \citenamefont {de~Tomasi},
  \citenamefont {Fioretti}, \citenamefont {Nosbaum}, \citenamefont {Akulin},\
  and\ \citenamefont {Pillet}}]{moco+:98}%
  \BibitemOpen
  \bibfield  {author} {\bibinfo {author} {\bibfnamefont {I.}~\bibnamefont
  {Mourachko}}, \bibinfo {author} {\bibfnamefont {D.}~\bibnamefont {Comparat}},
  \bibinfo {author} {\bibfnamefont {F.}~\bibnamefont {de~Tomasi}}, \bibinfo
  {author} {\bibfnamefont {A.}~\bibnamefont {Fioretti}}, \bibinfo {author}
  {\bibfnamefont {P.}~\bibnamefont {Nosbaum}}, \bibinfo {author} {\bibfnamefont
  {V.~M.}\ \bibnamefont {Akulin}}, \ and\ \bibinfo {author} {\bibfnamefont
  {P.}~\bibnamefont {Pillet}},\ }\href {\doibase 10.1103/PhysRevLett.80.253}
  {\bibfield  {journal} {\bibinfo  {journal} {Phys. Rev. Lett.}\ }\textbf
  {\bibinfo {volume} {80}},\ \bibinfo {pages} {253} (\bibinfo {year}
  {1998})}\BibitemShut {NoStop}%
\bibitem [{\citenamefont {Anderson}\ \emph {et~al.}(1998)\citenamefont
  {Anderson}, \citenamefont {Veale},\ and\ \citenamefont
  {Gallagher}}]{anve+:98}%
  \BibitemOpen
  \bibfield  {author} {\bibinfo {author} {\bibfnamefont {W.~R.}\ \bibnamefont
  {Anderson}}, \bibinfo {author} {\bibfnamefont {J.~R.}\ \bibnamefont {Veale}},
  \ and\ \bibinfo {author} {\bibfnamefont {T.~F.}\ \bibnamefont {Gallagher}},\
  }\href {\doibase 10.1103/PhysRevLett.80.249} {\bibfield  {journal} {\bibinfo
  {journal} {Phys. Rev. Lett.}\ }\textbf {\bibinfo {volume} {80}},\ \bibinfo
  {pages} {249} (\bibinfo {year} {1998})}\BibitemShut {NoStop}%
\bibitem [{\citenamefont {Li}\ \emph {et~al.}(2005)\citenamefont {Li},
  \citenamefont {Tanner},\ and\ \citenamefont {Gallagher}}]{lita+:05}%
  \BibitemOpen
  \bibfield  {author} {\bibinfo {author} {\bibfnamefont {W.}~\bibnamefont
  {Li}}, \bibinfo {author} {\bibfnamefont {P.~J.}\ \bibnamefont {Tanner}}, \
  and\ \bibinfo {author} {\bibfnamefont {T.~F.}\ \bibnamefont {Gallagher}},\
  }\href {\doibase 10.1103/PhysRevLett.94.173001} {\bibfield  {journal}
  {\bibinfo  {journal} {Phys. Rev. Lett.}\ }\textbf {\bibinfo {volume} {94}},\
  \bibinfo {pages} {173001} (\bibinfo {year} {2005})}\BibitemShut {NoStop}%
\bibitem [{\citenamefont {van Ditzhuijzen}\ \emph {et~al.}(2008)\citenamefont
  {van Ditzhuijzen}, \citenamefont {Koenderink}, \citenamefont {Hern\'{a}ndez},
  \citenamefont {Robicheaux}, \citenamefont {Noordam},\ and\ \citenamefont {van
  Linden van~den Heuvell}}]{diko+:08}%
  \BibitemOpen
  \bibfield  {author} {\bibinfo {author} {\bibfnamefont {C.~S.~E.}\
  \bibnamefont {van Ditzhuijzen}}, \bibinfo {author} {\bibfnamefont {A.~F.}\
  \bibnamefont {Koenderink}}, \bibinfo {author} {\bibfnamefont {J.~V.}\
  \bibnamefont {Hern\'{a}ndez}}, \bibinfo {author} {\bibfnamefont
  {F.}~\bibnamefont {Robicheaux}}, \bibinfo {author} {\bibfnamefont {L.~D.}\
  \bibnamefont {Noordam}}, \ and\ \bibinfo {author} {\bibfnamefont {H.~B.}\
  \bibnamefont {van Linden van~den Heuvell}},\ }\href {\doibase
  10.1103/PhysRevLett.100.243201} {\bibfield  {journal} {\bibinfo  {journal}
  {Phys. Rev. Lett.}\ }\textbf {\bibinfo {volume} {100}},\ \bibinfo {eid}
  {243201} (\bibinfo {year} {2008})}\BibitemShut {NoStop}%
\bibitem [{\citenamefont {Nipper}\ \emph {et~al.}(2012)\citenamefont {Nipper},
  \citenamefont {Balewski}, \citenamefont {Krupp}, \citenamefont {Butscher},
  \citenamefont {L\"ow},\ and\ \citenamefont {Pfau}}]{niba+:12}%
  \BibitemOpen
  \bibfield  {author} {\bibinfo {author} {\bibfnamefont {J.}~\bibnamefont
  {Nipper}}, \bibinfo {author} {\bibfnamefont {J.~B.}\ \bibnamefont
  {Balewski}}, \bibinfo {author} {\bibfnamefont {A.~T.}\ \bibnamefont {Krupp}},
  \bibinfo {author} {\bibfnamefont {B.}~\bibnamefont {Butscher}}, \bibinfo
  {author} {\bibfnamefont {R.}~\bibnamefont {L\"ow}}, \ and\ \bibinfo {author}
  {\bibfnamefont {T.}~\bibnamefont {Pfau}},\ }\href {\doibase
  10.1103/PhysRevLett.108.113001} {\bibfield  {journal} {\bibinfo  {journal}
  {Phys. Rev. Lett.}\ }\textbf {\bibinfo {volume} {108}},\ \bibinfo {pages}
  {113001} (\bibinfo {year} {2012})}\BibitemShut {NoStop}%
\bibitem [{\citenamefont {Maxwell}\ \emph {et~al.}(2013)\citenamefont
  {Maxwell}, \citenamefont {Szwer}, \citenamefont {Paredes-Barato},
  \citenamefont {Busche}, \citenamefont {Pritchard}, \citenamefont {Gauguet},
  \citenamefont {Weatherill}, \citenamefont {Jones},\ and\ \citenamefont
  {Adams}}]{masz+:13}%
  \BibitemOpen
  \bibfield  {author} {\bibinfo {author} {\bibfnamefont {D.}~\bibnamefont
  {Maxwell}}, \bibinfo {author} {\bibfnamefont {D.~J.}\ \bibnamefont {Szwer}},
  \bibinfo {author} {\bibfnamefont {D.}~\bibnamefont {Paredes-Barato}},
  \bibinfo {author} {\bibfnamefont {H.}~\bibnamefont {Busche}}, \bibinfo
  {author} {\bibfnamefont {J.~D.}\ \bibnamefont {Pritchard}}, \bibinfo {author}
  {\bibfnamefont {A.}~\bibnamefont {Gauguet}}, \bibinfo {author} {\bibfnamefont
  {K.~J.}\ \bibnamefont {Weatherill}}, \bibinfo {author} {\bibfnamefont
  {M.~P.~A.}\ \bibnamefont {Jones}}, \ and\ \bibinfo {author} {\bibfnamefont
  {C.~S.}\ \bibnamefont {Adams}},\ }\href {\doibase
  10.1103/PhysRevLett.110.103001} {\bibfield  {journal} {\bibinfo  {journal}
  {Phys. Rev. Lett.}\ }\textbf {\bibinfo {volume} {110}},\ \bibinfo {pages}
  {103001} (\bibinfo {year} {2013})}\BibitemShut {NoStop}%
\bibitem [{\citenamefont {Frasier}\ \emph {et~al.}(1999)\citenamefont
  {Frasier}, \citenamefont {Celli},\ and\ \citenamefont {Blum}}]{frce+:99}%
  \BibitemOpen
  \bibfield  {author} {\bibinfo {author} {\bibfnamefont {J.~S.}\ \bibnamefont
  {Frasier}}, \bibinfo {author} {\bibfnamefont {V.}~\bibnamefont {Celli}}, \
  and\ \bibinfo {author} {\bibfnamefont {T.}~\bibnamefont {Blum}},\ }\href
  {\doibase 10.1103/PhysRevA.59.4358} {\bibfield  {journal} {\bibinfo
  {journal} {Phys. Rev. A}\ }\textbf {\bibinfo {volume} {59}},\ \bibinfo
  {pages} {4358} (\bibinfo {year} {1999})}\BibitemShut {NoStop}%
\bibitem [{\citenamefont {Robicheaux}\ \emph {et~al.}(2004)\citenamefont
  {Robicheaux}, \citenamefont {Hern{\'a}ndez}, \citenamefont {Top{\c{c}}u},\
  and\ \citenamefont {Noordam}}]{rohe+:04}%
  \BibitemOpen
  \bibfield  {author} {\bibinfo {author} {\bibfnamefont {F.}~\bibnamefont
  {Robicheaux}}, \bibinfo {author} {\bibfnamefont {J.~V.}\ \bibnamefont
  {Hern{\'a}ndez}}, \bibinfo {author} {\bibfnamefont {T.}~\bibnamefont
  {Top{\c{c}}u}}, \ and\ \bibinfo {author} {\bibfnamefont {L.~D.}\ \bibnamefont
  {Noordam}},\ }\href {\doibase 10.1103/PhysRevA.70.042703} {\bibfield
  {journal} {\bibinfo  {journal} {Phys. Rev. A}\ }\textbf {\bibinfo {volume}
  {70}},\ \bibinfo {pages} {042703} (\bibinfo {year} {2004})}\BibitemShut
  {NoStop}%
\bibitem [{\citenamefont {M\"ulken}\ \emph {et~al.}(2007)\citenamefont
  {M\"ulken}, \citenamefont {Blumen}, \citenamefont {Amthor}, \citenamefont
  {Giese}, \citenamefont {Reetz-Lamour},\ and\ \citenamefont
  {Weidem\"uller}}]{mubl+:07}%
  \BibitemOpen
  \bibfield  {author} {\bibinfo {author} {\bibfnamefont {O.}~\bibnamefont
  {M\"ulken}}, \bibinfo {author} {\bibfnamefont {A.}~\bibnamefont {Blumen}},
  \bibinfo {author} {\bibfnamefont {T.}~\bibnamefont {Amthor}}, \bibinfo
  {author} {\bibfnamefont {C.}~\bibnamefont {Giese}}, \bibinfo {author}
  {\bibfnamefont {M.}~\bibnamefont {Reetz-Lamour}}, \ and\ \bibinfo {author}
  {\bibfnamefont {M.}~\bibnamefont {Weidem\"uller}},\ }\href@noop {} {\bibfield
   {journal} {\bibinfo  {journal} {Phys. Rev. Lett.}\ }\textbf {\bibinfo
  {volume} {99}},\ \bibinfo {pages} {090601} (\bibinfo {year}
  {2007})}\BibitemShut {NoStop}%
\bibitem [{\citenamefont {M\"{u}ller}\ \emph {et~al.}(2008)\citenamefont
  {M\"{u}ller}, \citenamefont {Liang}, \citenamefont {Lesanovsky},\ and\
  \citenamefont {Zoller}}]{muli+:08}%
  \BibitemOpen
  \bibfield  {author} {\bibinfo {author} {\bibfnamefont {M.}~\bibnamefont
  {M\"{u}ller}}, \bibinfo {author} {\bibfnamefont {L.}~\bibnamefont {Liang}},
  \bibinfo {author} {\bibfnamefont {I.}~\bibnamefont {Lesanovsky}}, \ and\
  \bibinfo {author} {\bibfnamefont {P.}~\bibnamefont {Zoller}},\ }\href
  {http://stacks.iop.org/1367-2630/10/093009} {\bibfield  {journal} {\bibinfo
  {journal} {New J. Phys.}\ }\textbf {\bibinfo {volume} {10}},\ \bibinfo
  {pages} {093009} (\bibinfo {year} {2008})}\BibitemShut {NoStop}%
\bibitem [{\citenamefont {W\"uster}\ \emph {et~al.}(2010)\citenamefont
  {W\"uster}, \citenamefont {Ates}, \citenamefont {Eisfeld},\ and\
  \citenamefont {Rost}}]{wuat+:10}%
  \BibitemOpen
  \bibfield  {author} {\bibinfo {author} {\bibfnamefont {S.}~\bibnamefont
  {W\"uster}}, \bibinfo {author} {\bibfnamefont {C.}~\bibnamefont {Ates}},
  \bibinfo {author} {\bibfnamefont {A.}~\bibnamefont {Eisfeld}}, \ and\
  \bibinfo {author} {\bibfnamefont {J.~M.}\ \bibnamefont {Rost}},\ }\href
  {\doibase 10.1103/PhysRevLett.105.053004} {\bibfield  {journal} {\bibinfo
  {journal} {Phys. Rev. Lett.}\ }\textbf {\bibinfo {volume} {105}},\ \bibinfo
  {pages} {053004} (\bibinfo {year} {2010})}\BibitemShut {NoStop}%
\bibitem [{\citenamefont {W\"uster}\ \emph {et~al.}(2011)\citenamefont
  {W\"uster}, \citenamefont {Ates}, \citenamefont {Eisfeld},\ and\
  \citenamefont {Rost}}]{wuat+:11}%
  \BibitemOpen
  \bibfield  {author} {\bibinfo {author} {\bibfnamefont {S.}~\bibnamefont
  {W\"uster}}, \bibinfo {author} {\bibfnamefont {C.}~\bibnamefont {Ates}},
  \bibinfo {author} {\bibfnamefont {A.}~\bibnamefont {Eisfeld}}, \ and\
  \bibinfo {author} {\bibfnamefont {J.~M.}\ \bibnamefont {Rost}},\ }\href
  {http://stacks.iop.org/1367-2630/13/i=7/a=073044} {\bibfield  {journal}
  {\bibinfo  {journal} {New Journal of Physics}\ }\textbf {\bibinfo {volume}
  {13}},\ \bibinfo {pages} {073044} (\bibinfo {year} {2011})}\BibitemShut
  {NoStop}%
\bibitem [{\citenamefont {Scholak}\ \emph {et~al.}(2011)\citenamefont
  {Scholak}, \citenamefont {Wellens},\ and\ \citenamefont
  {Buchleitner}}]{scwe+:11}%
  \BibitemOpen
  \bibfield  {author} {\bibinfo {author} {\bibfnamefont {T.}~\bibnamefont
  {Scholak}}, \bibinfo {author} {\bibfnamefont {T.}~\bibnamefont {Wellens}}, \
  and\ \bibinfo {author} {\bibfnamefont {A.}~\bibnamefont {Buchleitner}},\
  }\href {http://stacks.iop.org/0953-4075/44/i=18/a=184012} {\bibfield
  {journal} {\bibinfo  {journal} {Journal of Physics B: Atomic, Molecular and
  Optical Physics}\ }\textbf {\bibinfo {volume} {44}},\ \bibinfo {pages}
  {184012} (\bibinfo {year} {2011})}\BibitemShut {NoStop}%
\bibitem [{\citenamefont {Kiffner}\ \emph {et~al.}(2012)\citenamefont
  {Kiffner}, \citenamefont {Park}, \citenamefont {Li},\ and\ \citenamefont
  {Gallagher}}]{kipa+:12}%
  \BibitemOpen
  \bibfield  {author} {\bibinfo {author} {\bibfnamefont {M.}~\bibnamefont
  {Kiffner}}, \bibinfo {author} {\bibfnamefont {H.}~\bibnamefont {Park}},
  \bibinfo {author} {\bibfnamefont {W.}~\bibnamefont {Li}}, \ and\ \bibinfo
  {author} {\bibfnamefont {T.~F.}\ \bibnamefont {Gallagher}},\ }\href {\doibase
  10.1103/PhysRevA.86.031401} {\bibfield  {journal} {\bibinfo  {journal} {Phys.
  Rev. A}\ }\textbf {\bibinfo {volume} {86}},\ \bibinfo {pages} {031401}
  (\bibinfo {year} {2012})}\BibitemShut {NoStop}%
\bibitem [{\citenamefont {Bariani}\ \emph {et~al.}(2012)\citenamefont
  {Bariani}, \citenamefont {Dudin}, \citenamefont {Kennedy},\ and\
  \citenamefont {Kuzmich}}]{badu+:12}%
  \BibitemOpen
  \bibfield  {author} {\bibinfo {author} {\bibfnamefont {F.}~\bibnamefont
  {Bariani}}, \bibinfo {author} {\bibfnamefont {Y.~O.}\ \bibnamefont {Dudin}},
  \bibinfo {author} {\bibfnamefont {T.~A.~B.}\ \bibnamefont {Kennedy}}, \ and\
  \bibinfo {author} {\bibfnamefont {A.}~\bibnamefont {Kuzmich}},\ }\href
  {\doibase 10.1103/PhysRevLett.108.030501} {\bibfield  {journal} {\bibinfo
  {journal} {Phys. Rev. Lett.}\ }\textbf {\bibinfo {volume} {108}},\ \bibinfo
  {pages} {030501} (\bibinfo {year} {2012})}\BibitemShut {NoStop}%
\bibitem [{\citenamefont {Petrosyan}\ \emph {et~al.}(2013)\citenamefont
  {Petrosyan}, \citenamefont {H\"oning},\ and\ \citenamefont
  {Fleischhauer}}]{peho+:13}%
  \BibitemOpen
  \bibfield  {author} {\bibinfo {author} {\bibfnamefont {D.}~\bibnamefont
  {Petrosyan}}, \bibinfo {author} {\bibfnamefont {M.}~\bibnamefont {H\"oning}},
  \ and\ \bibinfo {author} {\bibfnamefont {M.}~\bibnamefont {Fleischhauer}},\
  }\href {\doibase 10.1103/PhysRevA.87.053414} {\bibfield  {journal} {\bibinfo
  {journal} {Phys. Rev. A}\ }\textbf {\bibinfo {volume} {87}},\ \bibinfo
  {pages} {053414} (\bibinfo {year} {2013})}\BibitemShut {NoStop}%
\bibitem [{\citenamefont {Lesanovsky}\ \emph {et~al.}(2010)\citenamefont
  {Lesanovsky}, \citenamefont {Olmos.},\ and\ \citenamefont
  {Garrahan}}]{leol+:10}%
  \BibitemOpen
  \bibfield  {author} {\bibinfo {author} {\bibfnamefont {I.}~\bibnamefont
  {Lesanovsky}}, \bibinfo {author} {\bibfnamefont {B.}~\bibnamefont {Olmos.}},
  \ and\ \bibinfo {author} {\bibfnamefont {J.~P.}\ \bibnamefont {Garrahan}},\
  }\href {\doibase 10.1103/PhysRevLett.105.100603} {\bibfield  {journal}
  {\bibinfo  {journal} {Phys. Rev. Lett.}\ }\textbf {\bibinfo {volume} {105}},\
  \bibinfo {pages} {100603} (\bibinfo {year} {2010})}\BibitemShut {NoStop}%
\bibitem [{\citenamefont {Ates}\ \emph {et~al.}(2012)\citenamefont {Ates},
  \citenamefont {Garrahan},\ and\ \citenamefont {Lesanovsky}}]{atga+:12}%
  \BibitemOpen
  \bibfield  {author} {\bibinfo {author} {\bibfnamefont {C.}~\bibnamefont
  {Ates}}, \bibinfo {author} {\bibfnamefont {J.~P.}\ \bibnamefont {Garrahan}},
  \ and\ \bibinfo {author} {\bibfnamefont {I.}~\bibnamefont {Lesanovsky}},\
  }\href {\doibase 10.1103/PhysRevLett.108.110603} {\bibfield  {journal}
  {\bibinfo  {journal} {Phys. Rev. Lett.}\ }\textbf {\bibinfo {volume} {108}},\
  \bibinfo {pages} {110603} (\bibinfo {year} {2012})}\BibitemShut {NoStop}%
\bibitem [{\citenamefont {Ji}\ \emph {et~al.}(2013)\citenamefont {Ji},
  \citenamefont {Ates}, \citenamefont {Garrahan},\ and\ \citenamefont
  {Lesanovsky}}]{jiat+:13}%
  \BibitemOpen
  \bibfield  {author} {\bibinfo {author} {\bibfnamefont {S.}~\bibnamefont
  {Ji}}, \bibinfo {author} {\bibfnamefont {C.}~\bibnamefont {Ates}}, \bibinfo
  {author} {\bibfnamefont {J.~P.}\ \bibnamefont {Garrahan}}, \ and\ \bibinfo
  {author} {\bibfnamefont {I.}~\bibnamefont {Lesanovsky}},\ }\href
  {http://stacks.iop.org/1742-5468/2013/i=02/a=P02005} {\bibfield  {journal}
  {\bibinfo  {journal} {Journal of Statistical Mechanics: Theory and
  Experiment}\ }\textbf {\bibinfo {volume} {2013}},\ \bibinfo {pages} {P02005}
  (\bibinfo {year} {2013})}\BibitemShut {NoStop}%
\end{thebibliography}
\end{document}